\documentclass[ajp]{revtex4-1} 
\usepackage{amsmath}  % needed for \tfrac, \bmatrix, etc.
\usepackage{amsfonts} % needed for bold Greek, Fraktur, and blackboard bold
\usepackage{graphicx} % needed for figures

\begin{document}

% Be sure to use the \title, \author, \affiliation, and \abstract macros
% to format your title page.  Don't use lower-level macros to  manually
% adjust the fonts and centering.

\title{A Software-Based Lock-In Measurement for Student Laboratories}
% In a long title you can use \\ to force a line break at a certain location.

\author{David T. Chuss}
\email{david.chuss@villanova.edu} % optional
%\altaffiliation[permanent address: ]{800 E. Lancaster Ave., Villanova, PA 19085} % optional second address
% If there were a second author at the same address, we would put another 
% \author{} statement here.  Don't combine multiple authors in a single
% \author statement.
\affiliation{Department of Physics, Villanova University, 800 E. Lancaster Ave., Villanova, PA 19085}
% Please provide a full mailing address here.

% See the REVTeX documentation for more examples of author and affiliation lists.

\date{\today}

\begin{abstract}
A student laboratory experiment is presented that introduces the concept of a lock-in measurement
through the exploration of the relationship between the power detected from a modulated light source and the distance between the source and the detector. The lock-in measurement is done in software using time streams for both the reference and the detector signals. The achievable experimental sensitivity is shown to increase several orders of magnitude over the simple case without using modulation. In addition, the student becomes familiar with computational techniques pertinent to large data sets. 
\end{abstract}
% AJP requires an abstract for all regular article submissions.
% Abstracts are optional for submissions to the "Notes and Discussions" section.

\maketitle % title page is now complete

\section{Introduction} % Section titles are automatically converted to all-caps.
% Section numbering is automatic.

For many measurements in the physical sciences, the targeted signal is much smaller than both the background and the fluctuations in the total power on the detector (noise). The background must be measured and subtracted in these cases. In addition, the total system noise needs to be limited or mitigated by appropriate filtering. 
Environmental variables, which include temperature and opacity changes, variations in the background, electronic pickup, and thermalization times all contribute to noise that has a spectrum that can be expressed functionally in the frequency domain as  $1/f^\alpha$, where $\alpha$ is a positive integer. From the perspective of the time domain, this means that on longer time scales, the system is less stable. 

A powerful technique for mitigating time-varying backgrounds in measuring small signals is that of lock-in measurement\cite{dicke}.  For a small signal, it is often possible to modulate the source at a given frequency or set of frequencies.  The signal that is the target of the measurement is encoded at a frequency that is higher than those at which the $1/f$ contribution is dominant in the system.  The time stream data can then be demodulated (since one
knows the frequency and phase of the modulation). This demodulation process selects the component of the time stream that corresponds to the modulation frequency and phase of the source signal and rejects the rest of the background. 
The importance of introducing lock-in techniques in advanced undergraduate laboratories has been emphasized in previous work \cite{temple, wolfson, schofield, libbrecht}, due to its wide applicability in various fields of physics including condensed matter and astrophysics. 

This work describes a laboratory experiment that is used to introduce the concept of a lock-in measurement to undergraduate students.  The experiment utilizes a ``digital'' lock-in technique in which the data from the detector and the reference signal are both stored for software demodulation. This is advantageous for three reasons. First, lock-in amplifiers can be expensive, and this technique avoids this cost. Second, the code for demodulating the signal can be written by the student, providing an opportunity for the student to understand the details of the technique in a tangible way. Finally, in practice, lock-in measurements are commonly realized in software in applications beyond the laboratory\cite{weinreb}. A key feature here is that this technique provides the power of a lock-in measurement while still retaining the raw time-stream data. Because the data can be re-analyzed multiple times, potentially with different analysis parameters, this technique can provide additional information over hardware-implemented lock-in measurements. This information can be used to quantify systematic effects for subsequent removal.  In addition, multiple layers of modulation can be accommodated. For example, in cosmic microwave background measurements, the mapping of linear polarization over the sky is desired to search for direct evidence of an inflationary epoch.  These maps can be considered as compressions of the time stream in which a combination of reference signals are correlated with the signal\cite{kusaka}.  These can include a combination of pointing, instrument rotation, polarization modulator state, or other time-dependent modulations. In this case, the reference signals correspond to telemetry information that is time-synched with the data stream. 

In such research applications, consideration of the frequency domain is often essential for understanding the data. The experiment described here provides the student the opportunity to examine the data from the perspective of the frequency domain, which is essential for understanding the technique and consequently, the final data products. This importance has been emphasized in hardware implementations of lock-in amplifiers \cite{schofield}. The software-based technique described here provides a complementary approach, in which the frequency domain understanding can be conveyed using a single stream of data.

\subsection{Lock-In Theory}
A lock-in measurement requires two time streams of data: the time-ordered data that contain the target signal (plus noise) and the reference that conveys the time-dependence of the modulation.  The time-ordered data, $D(t)$, is the sum of a the unmodulated background, $N(t)$, and the source signal, $S_{\mathit{s}}$, multiplied by the modulation function, $\Psi_\mathit{mod}(t$). Note that the source needs to be constant on time scales much larger than the modulation time scale ($T$).
\begin{equation}
D(t)=N(t)+S_\mathit{s}\Psi_\mathit{mod}(t)
\end{equation}
Each of these components can be written as their inverse Fourier transforms.  
\begin{equation}
D(t)=\int^\infty_0[N(f)+S_s\Psi_\mathit{mod}(f)]e^{-2\pi i ft} df
\label{eq:sig}
\end{equation}
Here, $N(f)$ is the spectrum of the background and $S_s\Psi_\mathit{mod}(f)$ is the spectrum of the modulated source signal, and is therefore highly peaked at the modulation frequency (or frequencies) $f=f_m$ and approximately zero everywhere else. 

A lock-in measurement can be expressed as an integral of the product of the time stream and the reference signal. The new (demodulated) time stream is
\begin{equation}
S(t, t_0)=\frac{1}{T}\int_{t}^{t+T}D(t^\prime)R(t^\prime+t_0)dt^\prime,
\label{eq:lock}
\end{equation}
where $R(t)$ is the reference signal that describes the functional form of the modulation, and $t_0$ is and offset corresponding to a phase delay between the reference and detector signals. Note that this expression is the cross-correlation between the detector and reference signal. The value of the cross-correlation at an ``optimal'' value of $t_0$ will be the desired demodulated signal.

To simplify the argument, it will be assumed that the reference signal has a single Fourier component. However, in reality, a more general modulation function can be treated as a linear superposition of multiple Fourier components, so no generality is lost by this assumption. Letting $R(t,t_0)=e^{2\pi i f_mt+i\phi_m}$, where $f_m$ is the frequency corresponding to the modulation, and $\phi_m=2\pi f_m t_0$ is the phase between the reference signal and the detector response, 
\begin{equation}
S(t, t_0)=\frac{1}{T}\int_{t}^{t+T}D(t^\prime)e^{2\pi i f_mt^\prime+2\pi i f_mt_0}dt^\prime.
\label{eq:lock}
\end{equation}
The parameter $T\gg 1/f_m$ is the integration time and is generally a free parameter of the analysis; however, it must be sufficiently large that many modulation cycles occur. Expanding this using Equation~\ref{eq:sig} leads to the following expression. 
\begin{align}
S(t,t_0)&=\frac{1}{T}\int_{t}^{t+T}\left\{\int_0^\infty [N(f)+S_s\Psi_\mathit{mod}(f)]e^{-2\pi if t^\prime} df\right\}e^{2\pi if_mt^\prime+2\pi i f_mt_0}dt^\prime
\end{align}
Exchanging the order of the integration leads to 
\begin{align}
S(t,t_0)&=e^{2\pi i f_mt_0}\int_0^\infty[N(f)+S_s\Psi_\mathit{mod}(f)]\left\{\frac{1}{T}\int_{t}^{t+T} e^{2\pi i(f_m-f)t^\prime} dt^\prime\right\}df.
\end{align}
The integral within the braces is $\delta(f_m-f)/T$ when the integration extends over $\pm\infty$.  If $T$ is large enough, this is a good approximation. The only effect of a finite integration limits is that instead of a (zero width) Dirac delta function, the function will have a finite width and amplitude, but still will be strongly peaked at $f_m$.
We denote this function $\Phi_T(f,t)$ and also now define $\Delta f_m=1/T$ as the ``bandwidth'' of the observation.

\begin{equation}
S(t,t_0)=e^{2\pi i f_mt_0} \Delta f_m\int_0^{\infty}[N(f)+S_s\Psi_\mathit{mod}(f)]\Phi_T(f,t)df\
\label{eq:lockin}
\end{equation}
Recall that the $S_s\Psi_\mathit{mod}(f)$ is highly peaked around $f_m$ because the desired signal is modulated at this frequency. The same is true of $\Phi_T(f,t)$, because it is the modulation function.  Thus, the second term will be just the signal times a calibration constant. 
For the first term, ${N}(f)$ is a function with support at all frequencies, so the ``windowing'' applied by multiplying by $\Phi_T(f,t)$ and integrating will limit the background contribution to a small band around $f_m$. This means that constant backgrounds ($f=0$) are rejected, along with any stray signals at other frequencies or phases. 
In addition, because the background at $f_m$ are not phase locked to the reference, they will not contribute significantly to $S(t)$. In general, the background will be time dependent, since there will be a variance in $N(f)$. This is the fundamental noise of the system, which integrates to zero as the integration time increases. With the definition in Equation~\ref{eq:lock}, it can be seen that
\begin{equation}
S(t,t_0)\propto S_\mathit{s}.
\end{equation}
The demodulated signal is proportional to the source signal, as desired. The phase, $\phi_m$ (or equivalently, $t_0$) is often determined empirically to maximize the proportionality constant. 

\section{Experimental Setup}
%---------------------------------------------------------------
\begin{figure}[htbp]
   \centering
   \includegraphics[width=4in]{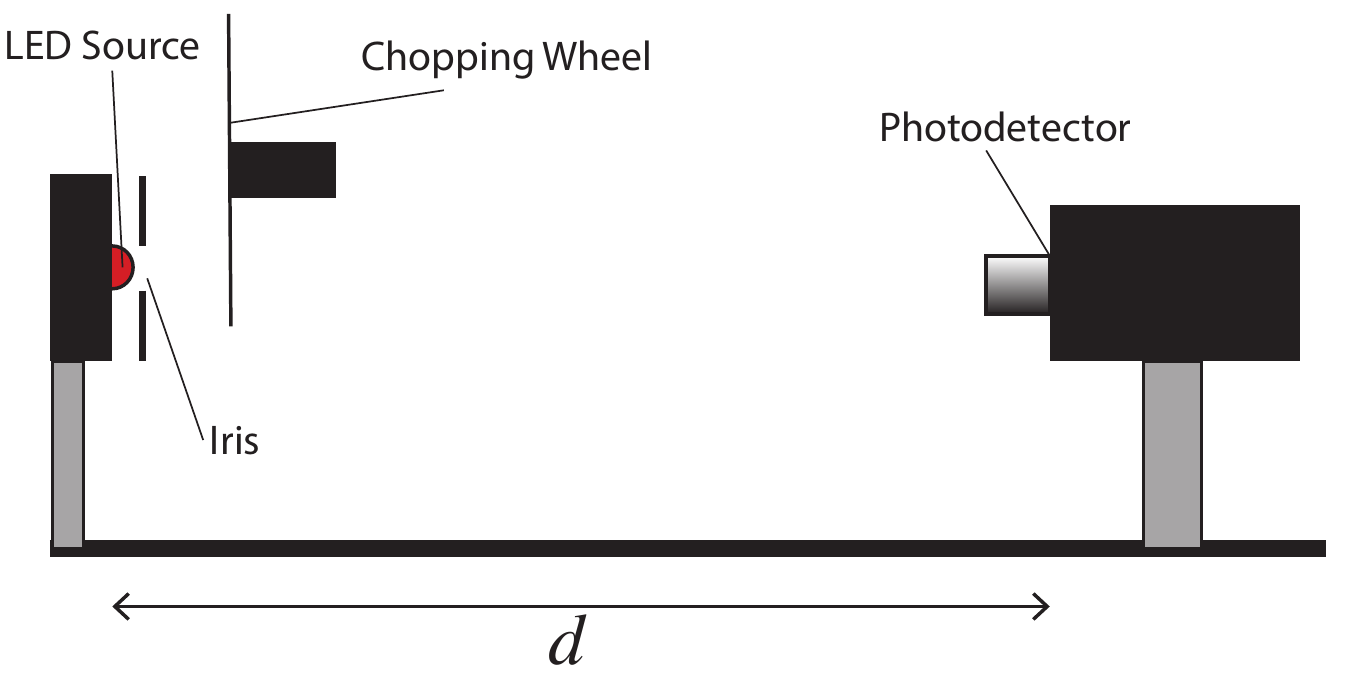}\\
   \includegraphics[width=4in, angle=180]{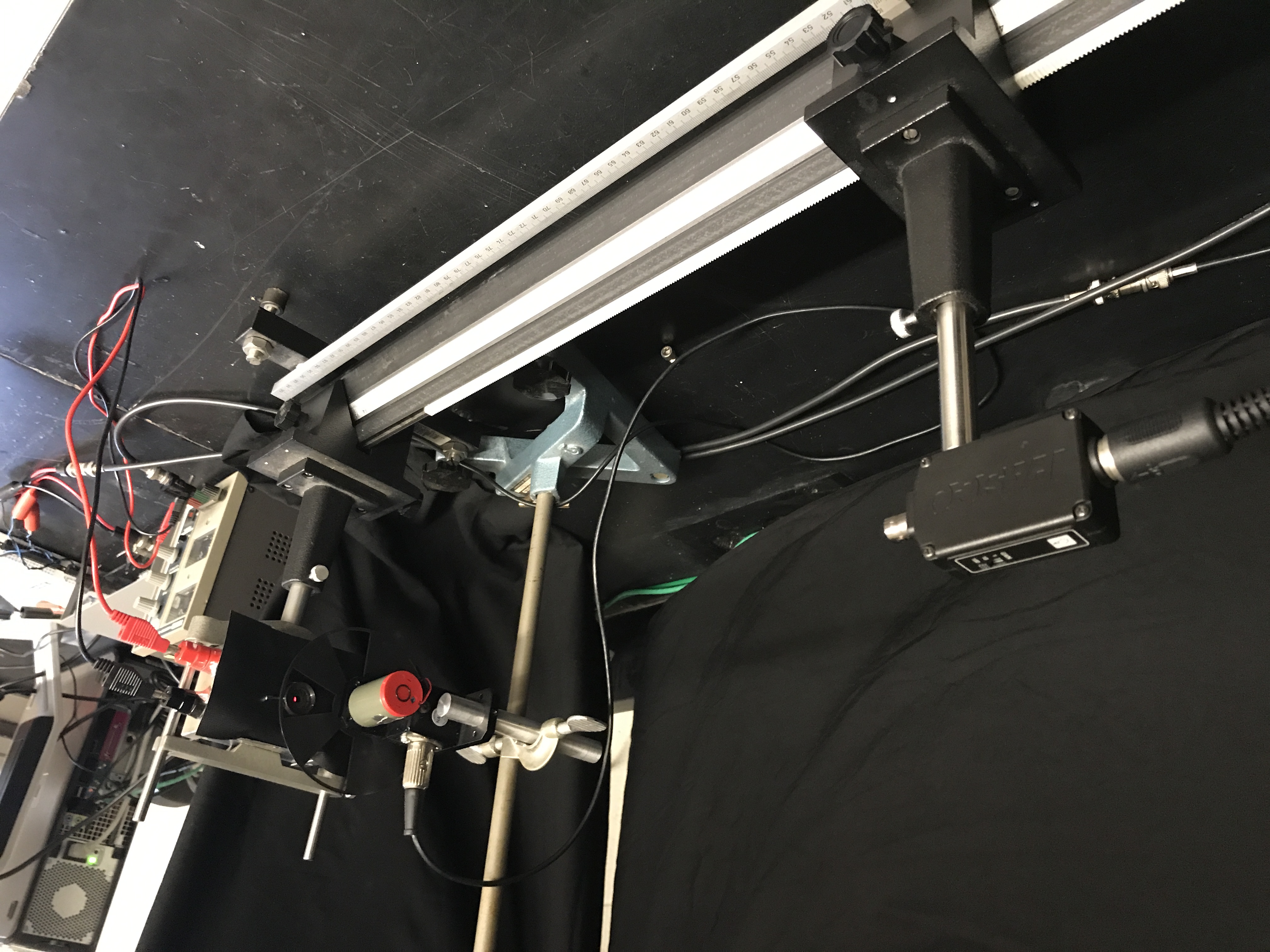} % requires the graphicx package
   \caption{(Top) The setup for the experiment is shown. An LED provides the source and is modulated by a chopper. The detector is placed a variable distance $d$ from the source. (Bottom) A photograph of the setup is shown. }
   \label{fig:setup}
\end{figure}
%---------------------------------------------------------------
The setup for the experiment is shown in Figure~\ref{fig:setup}. A red LED serves as the source. An iris is placed in front of the LED to provide an adjustable aperture. This reduces the effects of the LED lens in the near field. An optical chopper wheel is placed in front of the source to alternately block and reveal the LED to the detector. The detector uses is a PASCO CI-6504A Light Sensor and is used at its 100$\times$ gain setting. This setting corresponds to a maximum illuminance of 5 lux, which was found to be sufficient for measuring the LED plus the background ambient light.  The detector is mounted on a common rail with the source and aligned such that its center is at the same height as the LED.  During the course of the experiment, the distance, $d$, between the detector and the source is varied, and the detector signal is measured.

The analog output from the sensor along with the reference signal from the chopper are connected to analog inputs on an National Instruments USB-6210 interface.  A simple LabView VI is used to acquire time streams from the signal and reference at each distance separating the detector and the LED.  At each $d$, 20,000 samples are acquired at a rate of 600 Hz, corresponding to a total integration time of 33.3 s. The data taken at each $d$ are stored in a separate file.  Data sets are taken with the chopper on and off to demonstrate the difference between modulated and non-modulated techniques. 

\section{Data Analysis}
{\it Python} code is used to read in the files, examine the data, and perform the lock-in measurement. Depending on the skill level of the students, this code can either be supplied by the instructor for the use of the student, or generated by the students themselves from the principles of the lock-in technique.  

\begin{figure}[htbp]
   \centering
   \includegraphics[width=4in]{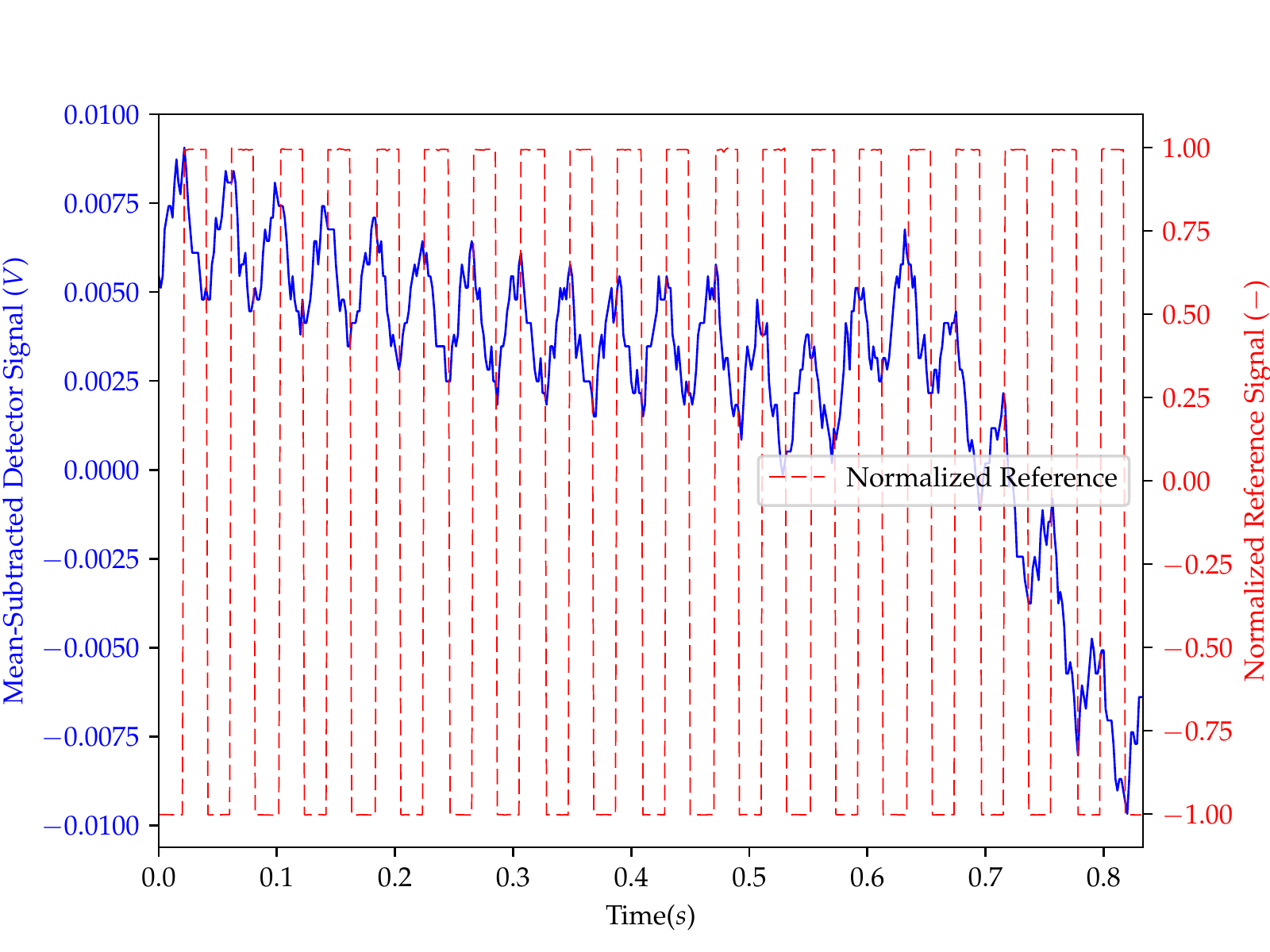}\\
   \includegraphics[width=4in]{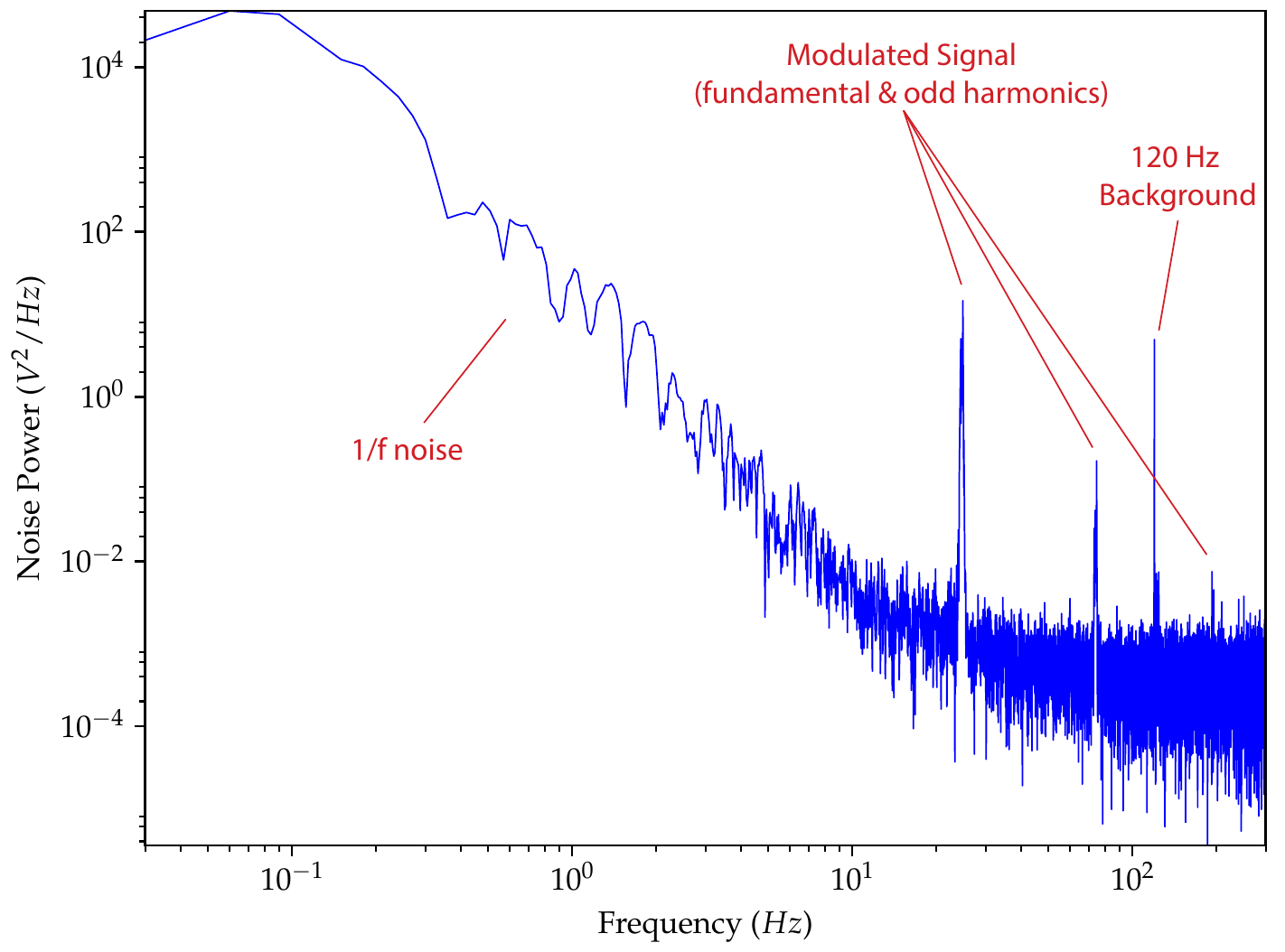} % requires the graphicx package
   \caption{(Top) A sample time stream is shown along with the concurrent reference signal for part of the signal corresponding to a distance of 165.7 cm. The mean of the signal over the time interval shown is subtracted for convenience of display. The baseline of the signal is seen to be variable in time.  (Bottom) The Fourier Transform for the same signal is shown. In this plot, the $1/f$ noise dominates at low frequencies. The frequencies associated with the modulated signal are indicated.}
   \label{fig:data}
\end{figure}

It is instructive for the student to visualize the data in both the time and frequency domains. In Figure~\ref{fig:data}, part of the time stream corresponding to $d=165.7$ cm is shown with its mean value subtracted for convenience. The normalized reference signal is overplotted. The modulated signal can be seen at the same frequency as the lock-in.  There is a phase offset between the signal and the reference that is primarily due to the location of the emitter-diode pair that measures the reference signal, though there could be electronic contributions as well.  Even on this very short ($<1$ s ) time scale, the signal is seen to be unstable, as the baseline drifts many times the level of the source (as measured by the difference between the peaks and valleys in the modulated signal).

The bottom plot of this figure shows the frequency spectrum of the same data. A Hanning filter was applied to the data before utilizing a Fast Fourier Transform (FFT). The instability on scales below 10 Hz is evident as the $1/f$ rise at low frequencies.   The modulation frequency is clearly seen at 24.7 Hz.  Its first two odd harmonics are also  visible, though most of the power in this signal is observed to be in the fundamental. This is primarily due to the bandwidth limitation of the detector.  {\it I.e.} the detector cannot respond with infinite bandwidth to the square wave modulation due to its limited temporal response, and so the detector acts as a low-pass filter.  The location of the modulated signal in frequency space clearly demonstrates the utility of the lock-in technique. That is, modulation moves the signal of interest to a frequency where the noise is orders of magnitude lower than those corresponding to the integration time of the detector.

There is also a spike at 120 Hz corresponding to the frequency of the fluorescent lights in the room. The modulation frequency was chosen such that its harmonics avoided this spectral feature. Due to the 600 Hz sampling frequency, the highest measured frequency (according to the Nyquist-Shannon sampling theorem) is 300 Hz.  Given the limited temporal response of the detector and the lack of an observed harmonic of the 120 Hz at 240 Hz, we anticipate that there is no significant effect of aliasing in our spectrum due to contributions beyond 300 Hz. If this experiment is reproduced in a noisier environment, or if a detector with larger bandwidth is used, a low-pass Nyquist filter can be placed in the signal path prior to digitization. 

%---------------------------------------------------------------
%---------------------------------------------------------------
\subsection{Demodulation}
The demodulation for this experiment is also done in the {\it Python} code and is a discretized version of the technique described above. There are two time streams: a reference ($R_n$) and a data signal ($D_n$).  The first step is to shift the reference relative to the data according to a phase, $\phi_m$.  Because the data are discrete, the arrays need to be shifted by an integral number of samples, corresponding to 
\begin{equation}
n_{\mathit{offset}}= \frac{\phi_m}{2\pi f_{\mathit{m}}t_{\mathit{samp}}},
\end{equation}
where $f_\mathit{m}$ is the modulation frequency, and $t_\mathit{samp}$ is the time for each sample. This number of samples is removed from the beginning of the $D_n$ array and at the end of the $R_n$ array. The optimal value of $\phi_m$ is determined by maximizing the signal-to-noise of the output signals, and a common value is used for all data points. 

Once the phase is taken into account, the time streams are divided into segments of length $T$. There are then $N=T/t_\mathit{samp}$ of these segments for each data file.
The $i$th signal value ($S_i$) in the demodulated time stream is then calculated via a discretized version of equation~\ref{eq:lock}.
\begin{equation}
S_i=\frac{t_\mathit{samp}}{T}\sum_{n=i}^{i+T/t_\mathit{samp}}R_nD_n
\end{equation}
The best estimate of the value of the signal at each distance is the mean of the $S_i's$, with the error corresponding to $\sqrt{\sigma^2/N}$, the standard deviation of the mean.

\subsection{Results}
\begin{figure}[htbp]
   \centering
   \includegraphics[width=4in]{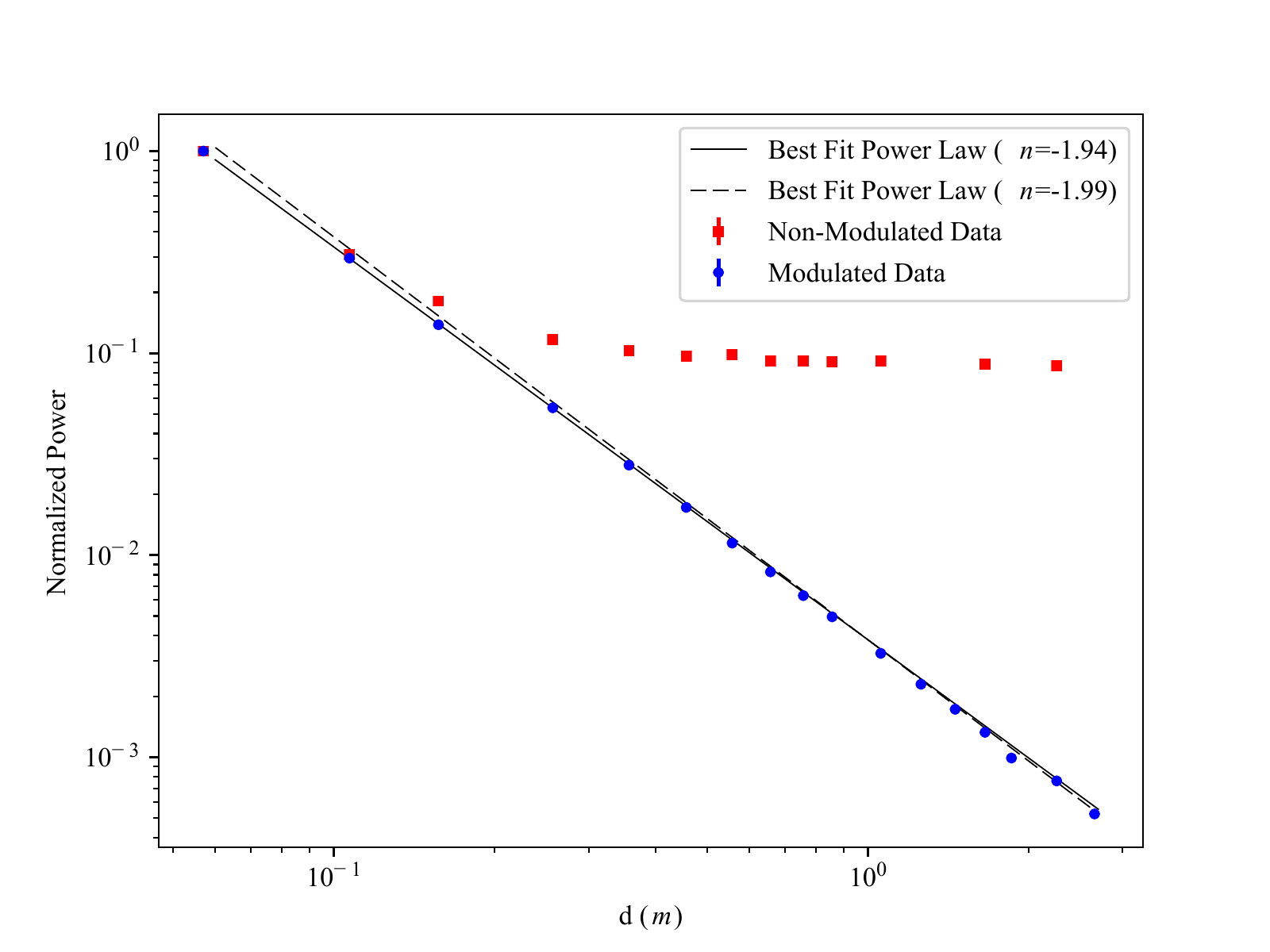}
   \caption{The detected power as a function of distance is shown for both modulated and unmodulated data.  In each case, the signal is normalized to the value at $d=5.7$ cm. In fitting the entire modulated data set, one obtains a power law index of $-1.94$.  In omitting the highest 5 points, the power law fit gives an index of $-1.99$.}
   \label{fig:results}
\end{figure}

Data were obtained at several distances between 5.7 cm and 270 cm. Results are shown in Figure~\ref{fig:results}. As a comparison, data taken without modulation are superposed on the plot. In this case, the mean of each 33.3 s time streams are plotted. Statistical uncertainties are approximated by the standard deviation of the mean. For both the modulated and unmodulated cases, the statistical errors are small compared to the size of the markers. For the modulated data, the point with the lowest signal-to-noise ratio ($SNR$) has $SNR>39$. Thus, residual variance are systematic in nature. One possible source of this error is stray light from reflections of the modulated source. Attempts were made to baffle potential light paths, and even at the largest distance examined, the source signal is anticipated to be larger than such reflections.  In examining the demodulated time streams, there are some anonymously low data points. Employing a deglitching alrogithm to the demodulated time streams (e.g. employing Chauvenet's criterion) provides a path for improving the systematic control. 

The data are fit to a simple power law ($y=\alpha x^\beta$). For the complete set of modulated data,  $\alpha=0.0038,\,\beta=-1.944$.  When excluding the highest five data points, $\alpha=0.0036,\,\beta=-1.994$.  For small values of $d$, the point source approximation begins to fail.  In the limit that the detector is infinitely close to the source, the illumination would be constant as a function of distance. At intermediate distances, one would expect functional dependence that is softer than an inverse-square law\cite{cataldo}. In the far-field, one anticipates the recovery of an inverse-square law.  In this work, the aperture diameter is 3.5 mm. At the closest distance (57 mm), the angular size of the source is $\sim 3.5^\circ$.  It is likely that the slightly lower value of the fitted power law index when all data are taken into account is a result of this geometric effect. This laboratory could easily be extended so that the student can explore near-field effects.  

%---------------------------------------------------------------
%---------------------------------------------------------------
 
\section{Summary}
An experiment to measure the power of a source on a detector as a function of distance has been described.  It is shown that by employing a lock-in technique, many orders of magnitude improvement in sensitivity can be obtained. This is mostly due to avoidance of the low-frequency noise associated with the environment.  The student laboratory described here offers practical insight into the nature of the lock-in technique by direct comparison with an unmodulated data set.  In addition, the software implementation of the demodulation provides experience with the algorithm without requiring an expensive lock-in amplifier. 

\begin{acknowledgments}

The author would like to thank Ed Wollack for helpful comments and discussion in preparing this manuscript.

\end{acknowledgments}

\end{document}